\documentclass[aps,prb,reprint,showpacs,superscriptaddress]{revtex4-2}
\usepackage{amsmath}
\usepackage{amssymb}
\usepackage{amsthm}
\usepackage{bbm}
\usepackage{graphicx}
\usepackage{hyperref}
\usepackage{physics}
\usepackage{systeme}
\usepackage{times}
\usepackage{epstopdf}
\usepackage{float}

\hypersetup{
  colorlinks=true,linkcolor=blue,citecolor=blue,
  filecolor=blue,urlcolor=blue,breaklinks=true
}

\begin{document}

\title{
  Modifications of electron states, magnetization and persistent current
  in a quantum dot by controlled curvature}

\author{Luis Fernando C. Pereira}
\email{luisfernandofisica@hotmail.com}
\affiliation{
        Departamento de F\'{i}sica,
        Universidade Federal do Maranh\~{a}o,
        65085-580, S\~{a}o Lu\'{i}s-MA, Brazil
      }

\author{Fabiano M. Andrade}
\email{fmandrade@uepg.br}
\affiliation{
        Departamento de Matem\'{a}tica e Estat\'{i}stica,
        Universidade Estadual de Ponta Grossa,
        84030-900 Ponta Grossa-PR, Brazil
      }

\author{Cleverson Filgueiras}
\email{cleverson.filgueiras@dfi.ufla.br}
\affiliation{
        Departamento de F\'{i}sica,
        Universidade Federal de Lavras, Caixa Postal 3037,
        37200-000, Lavras-MG, Brazil
      }

\author{Edilberto O. Silva}
\email{edilberto.silva@ufma.br}
\affiliation{
        Departamento de F\'{i}sica,
        Universidade Federal do Maranh\~{a}o,
        65085-580, S\~{a}o Lu\'{i}s-MA, Brazil
      }

\date{\today }

\begin{abstract}
In this work, we use the thin-layer quantization procedure to
study the physical implications due to curvature effects on a quantum
dot in the presence of an external magnetic field.
Among the various physical implications due to the curvature of the
system, we can mention the absence of the $m=0$ state is the most
relevant one.
The absence of it affects the Fermi energy and consequently the
thermodynamic properties of the system.
In the absence of magnetic fields,  we verify that the rotational
symmetry in the lateral confinement is preserved in the electronic
states of the system and its degeneracy with respect to the harmonicity
of the confining potential is broken.
In the presence of a magnetic field, however, the energies of the
electronic states in a quantum dot with a curvature are greater than
those  obtained for a quantum dot in a flat space, and the profile of
degeneracy changes when the field is varied.
We show that the curvature of the surface modifies the number of
subbands occupied in the Fermi energy. In the study of both magnetization and persistent currents, we observe that Aharonov-Bohm-type (AB-type) oscillations are present, whereas de Haas-van Alphen-type (dHvA) oscillations are not well defined.
\end{abstract}

\maketitle

\section{Introduction}
\label{intro}

Quantum dots are simply connected systems in which a two-dimensional
electron gas (2DEG) that is free to move on a flat surface is confined
laterally by a potential acting in all directions of the surface
\cite{SSR.2001.41.1}.
They are also known as artificial atoms where the
lateral confining potential replaces the potential of the
nucleus \cite{PRL.1990.65.108}.
Such confining potential may be, for instance, yielded by a
hard-wall or even by an harmonic oscillator-type parabolic
potential.
The energy spectrum then is fully discrete and it can be
studied by experiments of transport phenomena if the dot is weakly
coupled to wide 2DEG regions by tunnel barriers.
At zero temperature, the free energy is just the total energy of the
system.

The application of a magnetic field perpendicular to the surface of a quantum dot redefines the state of free energy.
Then, thermodynamic properties of the system as, for example, the
  magnetization, can be calculated \cite{Book.Landau.Statistical}.
Another property that results of the application of a magnetic
field is the persistent current.
However, we must remember that persistent currents were originally
defined in a quantum ring as a result of the Aharonov-Bohm (AB) flux
through the hole \cite{PRL.7.46.1961, PLA.1983.96.365}, which modifies
the boundary conditions of the wave function.
Consequently, all properties of the system are periodic functions of the
AB flux. Therefore, for a multiply connected geometry, we can calculate the persistent current using the Byers-Yang relation \cite{PRL.7.46.1961}. For a quantum dot it is not a problem, although it is a simply connected structure, as long as  the  wave  functions states are zero in the $r=0$ region. If this does not occur, we can calculate the persistent currents
using  the definition of the current density operator, as done in
\cite{PA.1993.200.504}, where the authors investigated the persistent
currents of a quantum dot in the strong magnetic field
regime.
An alternative procedure to calculate this quantity
at the quantum dot was accomplished in \cite{PRB.1999.60.5626} as a
limiting case of that obtained for the quantum
ring, and subsequently recovering the result found in
\cite{PA.1993.200.504}.

A 2DEG does not necessarily have to be a planar system.
Technological developments have made it possible to fabricate nano-objects
of various  shapes \cite{PE.2000.6.828,Nano.2001.12.399}.
In \cite{JACS.2005.127.13782} the synthesis of semiconductor
nanocones with a controllable apex angle was described.
These achievements have attracted great interest, both in the
experimental and theoretical points of view, when a 2DEG is held
in the presence of external fields
\cite{APL.2006.88.212113,PRB.2007.75.205309}.
The theoretical models among with the recent advances in the
manipulation of nanostructures allow the fabrication of quantum
systems in geometries with nontrivial topologies
\cite{PRB.2017.95.205426,PRB.2016.94.205125,PRB.2015.91.235308,
  PRB.1994.49.5097,PRL.2000.84.2223,EPL.2007.79.57001}.
Then, it is crucial to understand how the topology modifies the physical
properties of these nanodevices.
From the theoretical point of view, the physical implications exhibited
by the quantum system confined on a curved surface are of great interest.
When a quantum particle is strongly constrained to move on a curved surface, it
experiences an effective potential energy whose magnitude depends on the
local curvatures along the surface
\cite{AoP.1971.63.586,PRA.1981.23.1982,PRA.1982.25.2893}.
The approach employed to address such a system
follows the well-known thin-layer quantization procedure.
Several other contributions have been accomplished in the context of constrained particle in quantum mechanics using
this procedure in the last few years
\cite{PRB.2004.69.195313,PRA.2018.98.062112,PRA.2017.96.022116,PRA.2014.90.042117,AoP.2016.364.68,PRL.2008.100.230403,PRB.2013.87.174413,PRL.2014.112.257203,PRL.2010.105.206601,JMP.2019.60.023502,EJP.2016.38.015405,NTch.2016.27.135302,PE.2019.106.200,PRB.2018.97.241103}.
The main result described in such approach is that, even in
the absence of interactions of any nature, the electrons cannot move
around freely on the surface.
This implies that we can investigate mesoscopic physical systems
in curved geometries by simply controlling the local geometric curvature
of the surface and then accessing the physical properties of interest.

The purpose of this paper is to investigate the physical
implications caused by curvature effects on a quantum dot in the
presence of external magnetic fields.
We build our model taking into account the thin-layer quantization
procedure  and obtain the energy spectrum, the Fermi energy, the magnetization,
and the persistent current.
We discuss in detail the effects of the surface
curvature on such properties.
The system considered consists of noninteracting spinless
electrons in a quantum dot constrained to a conical
surface with an AB flux piercing through the center of the
conical surface. The system is subject to a constant magnetic field in the
$z$-direction as well.

\section{Description of the model}
\label{sec:model}

We are interested here in studying the motion of a spinless charged
particle constrained to move on a curved surface in the presence of
magnetic fields.
We employ the procedure of \cite{PRL.2008.100.230403} for studying
the quantum mechanics of a constrained particle, which is based on
da Costa's thin-layer quantization procedure  \cite{PRA.1981.23.1982}.
In  \cite{PRL.2008.100.230403},
by making a proper choice of the gauge, it was shown that the surface
and transverse dynamics are exactly separable.
In the transverse motion, the dynamics is described by a
one-dimensional Schr\"{o}dinger equation with a transverse potential,
whereas the motion on the surface is described by a two-dimensional
Schr\"odinger equation in which appears a geometric potential,
given in terms of the mean and the Gaussian curvatures.
The geometric potential is a consequence of the two-dimensional
confinement on the surface.

Let us consider, therefore, a non-interacting 2DEG constrained to
move on a curved surface in the presence of both a magnetic field
and a radial potential \cite{SST.1996.11.1635} given by
\begin{equation}
V\left( r\right) =\frac{a_{1}}{r^{2}}+a_{2}r^{2}-V_{0},  \label{pot.radial}
\end{equation}
with $V_{0}=2\sqrt{a_{1}a_{2}}$.
This radial potential has a minimum at
$r_{0}=\left( a_{1}/a_{2}\right) ^{1/4}$.
For $r\rightarrow r_{0}$, we obtain the parabolic potential model,
$V(r) \approx \mu \omega_{0}^{2}(r-r_{0})^{2}/2$, where
$\omega_{0}=\sqrt{8a_{2}/\mu }$ characterizes the strength
of the transverse confinement.
The potential (\ref{pot.radial}) describes a 2D quantum ring,
nevertheless, it can describe others physical systems.
For example, if $a_{1}=0$, we have a quantum dot and if $a_{2}=0$, we
have a quantum anti-dot.
Both the radius and the width of the ring can be adjusted independently
by suitably choosing $a_{1}$ and $a_{2}$.

In this work, the curved surface is defined by the following line element
in polar coordinates  \cite{AoP.2008.323.3150,JMP.2012.53.122106}
\begin{equation}
  \label{eq:line_element}
  ds^{2}=dr^{2}+\alpha ^{2}r^{2}d\theta ^{2},
\end{equation}
with $r\geq 0$ and $0\leq \theta <2\pi $. For $0<\alpha <1$ (deficit
angle), the metric above describes an actual conical surface,
while for $\alpha >1$ (proficit angle), it represents a saddle-like
surface.
In what follows we focus our analysis in a conical surface, in which
$0 < \alpha \leq 1$.
In this case, the Gaussian and the mean curvatures are
given, respectively, by \cite{EPL.2007.80.46002}
\begin{equation}
  \mathcal{K}=
  \left( \frac{1-\alpha }{\alpha }\right) \frac{\delta (r)}{r},
  \qquad
  \mathcal{H}=\frac{\sqrt{1-\alpha ^{2}}}{2\alpha r}.  \label{curvat}
\end{equation}
and the corresponding geometric potential is written as
\begin{equation}
V_{g}(r)=-\frac{\hbar ^{2}}{2\mu}\left[ \frac{(1-\alpha ^{2})}{4\alpha
^{2}r^{2}}-\left( \frac{1-\alpha }{\alpha }\right) \frac{\delta (r)}{r}
\right],  \label{Vgeo}
\end{equation}
with $\mu$ is the electron effective mass.
For the field configuration, we consider a superposition of magnetic
fields, $\mathbf{B}=\mathbf{B}_{1}+\mathbf{B}_{2}$, with
$\mathbf{B_{1}}=B\mathbf{\hat{z}}$ being a uniform magnetic field and
$\mathbf{B_{2}}=(l\hbar/e\alpha r)\delta(r)\mathbf{\hat{z}}$ being a
magnetic flux tube, with $l=\varPhi/\varPhi_{0}$ being the AB
flux parameter, $e$ is the electric charge, and $\varPhi_{0}=h/e$ is the
magnetic flux quantum.
The field $\mathbf{B}$ is obtained from the vector potential $\mathbf{A}=\mathbf{A}_{1}+\mathbf{A}_{2}$, where $\mathbf{A}_{1}=(Br/2\alpha)\boldsymbol{\hat{\varphi}}$ and $\mathbf{A}_{2}=(l\hbar/e\alpha r)\boldsymbol{\hat{\varphi}}$.

Since we are only interested in the dynamics on the surface, we ignore
the transverse one.
Thus, the relevant equation is
\begin{equation}
H\chi _{S}\left( r,\varphi \right) =E\chi_{S}\left( r,\varphi \right), \label{chr}
\end{equation}
where
\begin{align}
H = {} & -\frac{\hbar ^{2}}{2\mu }\left[ \frac{1}{r}\frac{\partial }{\partial r}
\left( r\frac{\partial }{\partial r}\right) +\frac{1}{\alpha ^{2}r^{2}}
\left( \frac{\partial }{\partial \varphi }-il\right) ^{2}\right]\notag  \\
&-\frac{\hbar ^{2}}{2\mu }\left[ \frac{ieB}{\hbar \alpha ^{2}}\left( \frac{
        \partial }{\partial \varphi }-il\right) +\frac{e^{2}B^{2}r^{2}}{4\hbar
        ^{2}\alpha ^{2}}\right]\notag  \\
&-\frac{\hbar ^{2}}{2\mu}\left[ \frac{(1-\alpha ^{2})}{4\alpha
        ^{2}r^{2}}-\left( \frac{1-\alpha }{\alpha }\right) \frac{\delta (r)}{r}
\right]\notag
\\
& +\frac{a_{1}}{r^{2}}+a_{2}r^{2}-V_{0}  \label{Hm}
\end{align}
is the Hamiltonian of the system.
Due to the presence of the singular $\delta$-function in
  $V_{g}(r)$, the Hamiltonian (\ref{Hm}) is not self-adjoint
  \cite{Book.2004.Albeverio}.
Consequently, the appropriated manner to solve the problem in
\eqref{chr} is by using the self-adjoint extension approach.
Thus,  to find the energy spectrum, we can use any of the methods
proposed in  \cite{PRL.1990.64.503} or \cite{PRD.2012.85.041701}, which
have been used to solve the spin-1/2 AB problem in curved space.
In this manner, the energy eigenvalues and wavefunctions of
Eq. (\ref{chr}) are given by
\begin{equation}
  E_{n,m}=
  \left( n+\frac{1}{2}\pm \frac{L}{2}\right) \hbar\omega
  - \frac{(m-l)}{2\alpha^{2}} \hbar \omega_{c}
  - \frac{\mu\omega_{0}^{2}r_{0}^{2}}{4},
  \label{Enm_dot}
\end{equation}
and
\begin{align}
  \chi _{S}\left( r,\varphi \right)  = {}
  & \left(\frac{1}{2\lambda ^{2}}\right)^{\frac{1+L}{2}}
    e^{im\varphi }e^{-\frac{r^2}{4 \lambda^2}} r^{L}  \nonumber \\
  & \times
    \Bigg[
    c_{1}\mathrm{M}
    \left(
    \frac{{1+L+{\lambda^{2}k^{2}}}}{2}{,\,1+L,\,\frac{r^2}{2\lambda^2}}
    \right)   \nonumber \\
  & + c_{2}{\mathrm{U}}
    \left(
    \frac{{1+L+{\lambda^{2}k^{2}}}}{2}{,\,1+L,\,\frac{r^2}{2\lambda^2}}
    \right)
    \Bigg], \label{solution}
\end{align}
where $n=0,1,2,\ldots$, $m=0,\pm 1, \pm 2, \ldots$, $\mathrm{M}$ and
$\mathrm{U}$ denote the confluent hypergeometric functions of the first
and the second kind, respectively \cite{Book.1972.Abramowitz}, and
$c_{i}$ ($i=1,2$) are constants.
The sign $+$ ($-$) in Eq. (\ref{Enm_dot}) refers to the
energy associate with the regular (irregular) solution for the
wavefunction at the origin. The effective angular momentum
\begin{equation}
  L=\sqrt{\frac{(m-l)^{2}}{\alpha ^{2}}+
    \frac{2a_{1}\mu }{\hbar ^{2}}-
    \frac{1-\alpha ^{2}}{4\alpha ^{2}}},
  \label{Moment}
\end{equation}
controls when the Hamiltonian is self adjoint:
if $|L|\geq 1$ it is self-adjoint and if $|L|<1$ it is not
self-adjoint.
The irregular solution must be taking into account only when it is not
self-adjoint (for more details, see Refs. \cite{PRD.2012.85.041701,AoP.2013.339.510}).
In Eq. (\ref{solution})
\begin{equation}
  k^{2}=
  \frac{2\mu E}{\hbar ^{2}}+
  \frac{\mu \omega }{\hbar\alpha ^{2}}\left( m+l\right) ,
  \label{k}
\end{equation}
is the wave number,
\begin{equation}
  \omega=\sqrt{\frac{\omega_{c}^{2}}{\alpha ^{2}}+\omega_{0}^{2}},
  \label{F}
\end{equation}
is the effective cyclotron frequency, and
\begin{equation}
  \lambda=\sqrt{\frac{\hbar}{\mu \omega},}
  \label{lambda}
\end{equation}
is the effective magnetic length, with $\omega_{c}=eB/\mu$ being the
usual cyclotron frequency.
The quantum number $n$ characterizes the radial
motion and it is viewed as the subband index, while the quantum
number $m$ characterizes the angular momentum.

\section{Analysis of the electronic states and the energy level
  structure}
\label{sec:analysis}

In this section, we investigate the energy spectrum of the quantum
dot model described in Sec. \ref{sec:model}.
It can be obtained from Eq. \eqref{Enm_dot} by setting $a_1=0$,
leading to the radial potential $V(r)=a_{2}r^{2}$.
The resulting quantum dot is then characterized by $r_0=0$ and
$a_2 =\mu\omega_0^2/8$, with $\hbar\omega_0=0.459$ meV
\cite{PRB.1999.60.5626}.
For the numerical simulations we use values obtained from a 2D GaAs
heterostructure: the effective mass is $\mu=0.067 m_e$, where $m_e$ is
the electron mass and the number of electrons is $N_e=1400$.

Returning to Eq. \eqref{Moment}, an analysis shows that in the case
of a quantum dot the state with $m=l$ is forbidden.
Here we consider only the effects of the constant magnetic field, then
$l=0$ and the state with $m=0$ is forbidden leading to $|L| \geq 1$.
Therefore, the Hamiltonian is self-adjoint for all other possible
values of $m$, and only the energy eigenvalues of the regular solution
($+$) need to be considered \cite{PRD.2012.85.041701}.
Moreover, this is a guarantee that the Gaussian curvature does not
modify the states of the dot and, therefore, we can neglect it.

We can immediately check that Eq. (\ref{Enm_dot}) gives the energy
eigenvalues for a quantum dot in a flat sample by setting $\alpha=1$
\cite{ZP.1928.47.446,PRL.1990.65.108,RPP.2001.64.701,SSR.2001.41.1,
RMP.2002.741283,Bird.2013}.
In the absence of an external magnetic field, we can verify that the
separation between the neighboring subbands is $\hbar\omega_{0}$ and it
is not influenced by the curvature of the surface.
On the other hand, in the presence
of it, we highlight two consequences:
(i) The separation between the neighboring subbands
is $\hbar \omega$, which depends on the magnetic field and $\alpha$ by
means of Eq. (\ref{F});
(ii) The energies of the states with $m>0$ are lower than those for
$m<0$.
The situation (i) implies that by either increasing the magnetic field,
or decreasing the value of $\alpha$, the distance between the subbands
increases, whereas the situation (ii) reveals that the eletrons tend to
fill the low-energy states with $m>0$.

Among the literature it is common to investigate the appearance of
degeneracy at a quantum dot.
For a quantum dot on a flat sample in the absence of an external
magnetic field, all the states are degenerates with exception of $m=0$:
the states with energies $E_{0,-1}$, $E_{0,1}$  form a ``shell" which is
two-fold degenerate;
the states with $E_{0,-2}, E_{0,2}, E_{1,0}$ form another ``shell" which
is three-fold degenerate;
and so on
(additional physical implications may occur, for example, if we consider
the spin of the electron, leading to the appearance of the magic numbers
$N=2,6,12,20,...$ \cite{RPP.2001.64.701,Bird.2013}).
The degeneracy with respect to the angular quantum number $m$ and $-m$
arises from the rotational symmetry in the lateral confinement while the
additional degeneracy, as for example $E_{0,-2}=E_{0,2}=E_{1,0}$, is
associated with the harmonicity of the confining potential
\cite{Bird.2013}.
In the presence of a magnetic field, though, the degeneracy is lifted
for small values of $B$, but as the magnetic field increases, accidental
degeneracies may occur, again leading to enhanced bunching of single-particle
levels \cite{RMP.2002.741283}.
In the case of a quantum dot on a curved sample in the
absence of a magnetic field, the energy becomes
\begin{equation}
E_{n,m}=\left( n+\frac{1}{2}+\frac{L}{2}\right) \hbar \omega_{0}.
\end{equation}
It is easy to see that the rotational symmetry in the lateral
confinement is preserved, but the degeneracy with respect to the
harmonicity of the confining potential is broken.
This result is a consequence of presence of both the mean
curvature (Eq. (\ref{curvat})) and disclination parameter $\alpha$.
Therefore, the degeneracy is reduced significantly.
On the other hand, with the presence of a magnetic field, the
profile of degeneracy changes when the field is varied.
In other words, we may have an increase or decrease of crossing of
levels depending on the value of the magnetic field.
At the limit when the transverse confinement tends to zero,
$\omega_{0}\rightarrow 0$, or, what is equivalent
$\omega_{c}\gg\omega_{0}$, we obtain
$E_{nm}=\left(N+1/2\right) \hbar\omega_{c}/\alpha$, with
$N=n+\left( L-m/\alpha \right) /2$.
Note that for $\alpha=1$, the usual Landau levels are recovered
\cite{PRL.1990.65.108}.
For $\alpha<1$, the curvature lifts the Landau Levels degeneracy.

\begin{figure}[!ht!]
  \centering
  \includegraphics[width=\columnwidth]{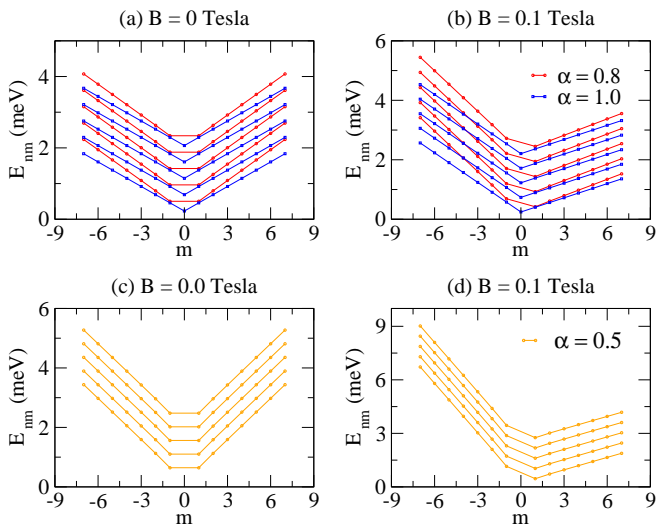}
  \caption{(Color online)
    The energy levels of a quantum dot as a function of the
    quantum number $m$ in the range of weak
    magnetic fields.
    We consider the first five subbands ($n=0,1,2,3,4$).
    In (a) and (b) are for $\alpha=0.8$ and $\alpha=1.0$, respectively.
    In (c) and (d), we consider $\alpha=0.5$.}
  \label{Subband}
\end{figure}

In Fig. \ref{Subband}, we plot the energies as a function of the quantum
number $m$ for three different values of $\alpha$ and two values
of magnetic field strength. The  circles and the squares represent the energies, while the solid lines indicate the subbands. For $B=0$, we can clearly see that the curve of the subbands shows a V-shaped format being more closed when there is curvature than
when it does not, that is, the energy of a ($n,m$) state of an electron is
larger at a quantum dot in a curved sample than in a flat one.
The electron states at a quantum dot in a flat sample are
degenerate with both rotational symmetries and harmonicity, whereas for
the curved case, degeneracy is only due to rotational symmetry.
We can see in Fig. \ref{Subband}(c) for $\alpha=0.5$ that the
curvature causes a small shift in the energies lifting
the degeneracy due to harmonicity.
The presence of a magnetic field breaks the rotational symmetry
and the subbands tend to undergo a clockwise rotation, in order to
have electrons that will occupy more states with positive energies
 (see Fig. \ref{Subband}(b) and Fig. \ref{Subband}(d)).
 This situation is independent of whether or not there is curvature.

\begin{figure}[ht!]
  \centering
  \includegraphics[width=\columnwidth]{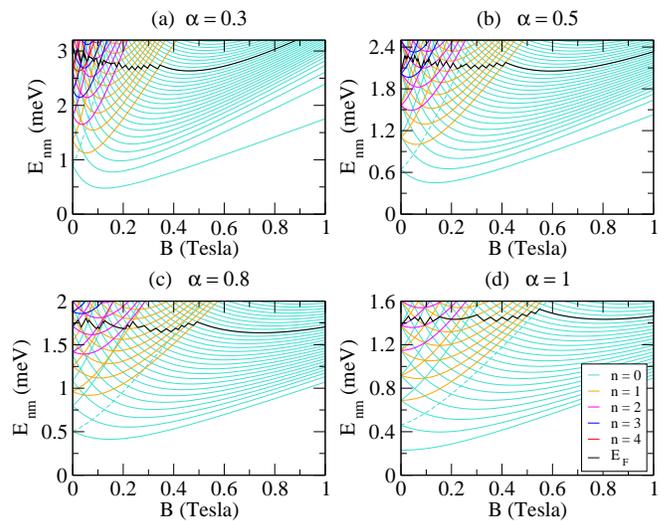}
  \caption{(Color online)
    The energy levels of a quantum dot as
    a function of the magnetic field.
    The black line represents the Fermi energy behavior for the case
    where there are 20 electrons in the sample.
    The dashed line corresponds to the energy $E_{0,1}$.}
  \label{Energy_all}
\end{figure}

In Fig. \ref{Energy_all} we show the behavior of the energy levels as a
function of the magnetic field for $\alpha=0.3$, $\alpha=0.5$,
$\alpha=0.8$ and $\alpha=1.0$.
We also plot the Fermi energy corresponding to the particular case where there are
$20$ electrons confined in a quantum dot.
In Fig. \ref{Energy_all}(d), it is possible see that when the
magnetic field $B\rightarrow 0^{+}$ Tesla the energy spectrum of a quantum dot is not
degenerated.
However, by increasing the intensity of the magnetic field, the
degeneracy appears in the spectrum and for some values of $B$ it might be quite high.
In Figs. \ref{Energy_all}(a), \ref{Energy_all}(b) and \ref{Energy_all}(c),
we can note that the crossing pattern of energy levels is affected
by the curvature.
We can observe that the states with $m<0$ are more impacted by the curvature as for example, the case of the energy $E_{0,-1}$ represented by
the dashed line. The behavior of the states with $m<0$ has important effects on the persistent current. Also, it is
remarkable in this figure that the amount of subbands occupied in the
Fermi energy is greater in the presence of curvature, although we also
notice that a subband is more quickly emptied.

\section{Magnetization}

\begin{figure}[!t!]
  \centering
  \includegraphics[width=\columnwidth]{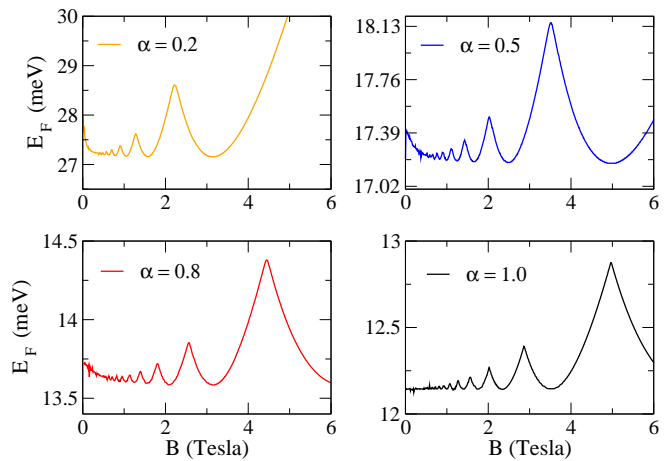}
  \caption{(Color online)
    The zero temperature Fermi energy as a function of magnetic
    field.}
  \label{Fermi_Energy_All}
\end{figure}

We start by studying the Fermi energy of the quantum dot discussed
in the previous section in a curved sample.
In Fig. \ref{Fermi_Energy_All}, we show the behavior of the Fermi
energy as a function of the magnetic field, and for some values of
$\alpha$.
We can observe that Fermi energy exhibits a non-smooth behavior when the
magnetic field is varied.
In the regime of weak magnetic fields, the Fermi energy exhibits a
downward deviation.
This is a direct consequence due to the absence of the $m=0$ state, and
hence the occupation of the states is altered; the minimum energy state
moves up one state.
A non-zero magnetic field makes the lowest-state energy to reach a
minimum value, as we can see from Figs. \ref{Energy_all}(b),
\ref{Energy_all}(c) and \ref{Energy_all}(d).
Once this minimum value has been reached, the absence of the $m=0$ state
has less effect on occupation.
Since it deals with electronic behavior, both the persistent current and
the magnetization will also have an anomalous result in the weak
magnetic field regime.

\begin{figure}[t!]
	\centering
	\includegraphics[width=\columnwidth]{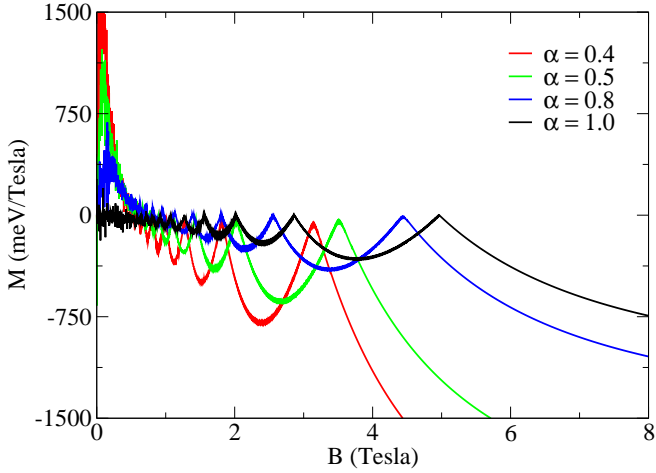}
	\caption{(Color online)
	The magnetization of the quantum dot as a function of magnetic field
	strength.}
	\label{Magnetizacao}
\end{figure}

\begin{figure}[h!]
	\centering
	\includegraphics[width=\columnwidth]{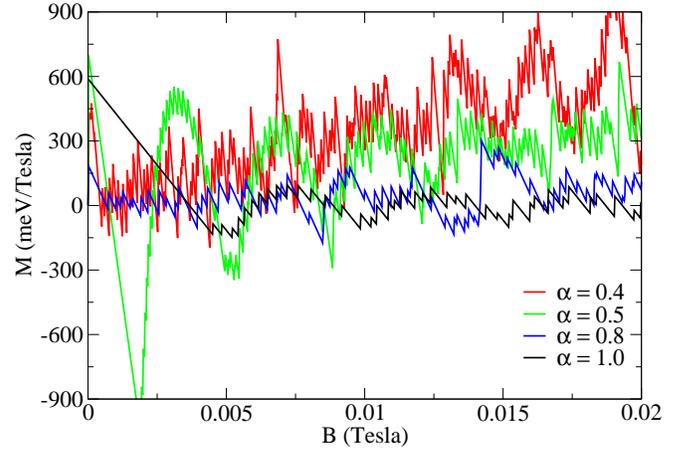}
	\caption{(Color online)
	The magnetization of the quantum dot in the weak magnetic fields interval.}
	\label{Magnetizacao-weak-B}\qquad
\end{figure}

\begin{figure}[h]
	\centering
	\includegraphics[width=\columnwidth]{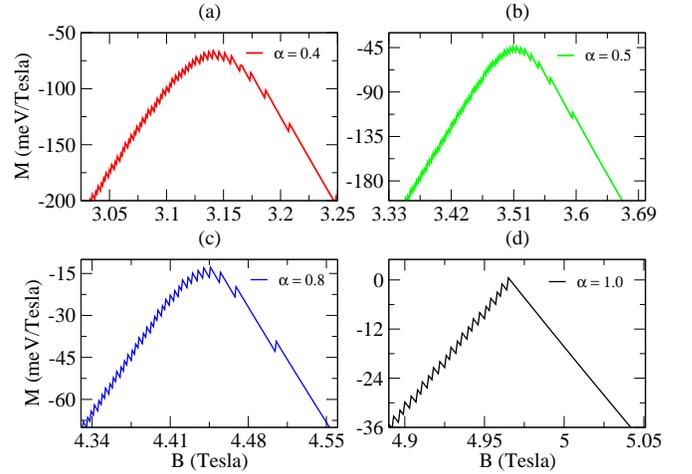}
	\caption{(Color online) Profile of the magnetization when the subband with $n=1$ is depopulated.}
	\label{Magnetization-1-subband}
\end{figure}

\begin{figure}[!t]
  \centering
  \includegraphics[width=\columnwidth]{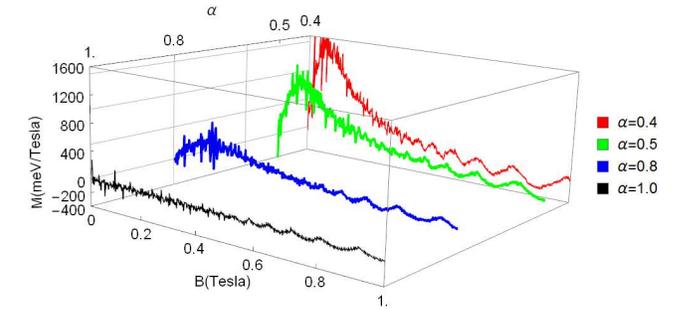}
  \caption{(Color online)
    The magnetization of the quantum dot as a function of magnetic field
    strength.
    The amplitude of the dHvA-type oscillations changes when the
    $\alpha$ parameter is changed.}
  \label{MagnetizationDot3d}
\end{figure}

The magnetization is a thermodynamic quantity which arises
as a response to the applied magnetic field.
At zero temperature, the free energy is just the total energy of system. Consequently, if the system is closed, the magnetization is given by
\begin{equation}
  M = -\frac{\partial U}{\partial B}
    = -\sum_{n,m}M_{n,m},  \label{Mag}
\end{equation}
where $M_{n,m}\equiv-\partial E_{n,m}/\partial B$ defines the
magnetic moment.
By using Eq. (\ref{Enm_dot}), the magnetic moment is written explicitly
as
\begin{equation}
  M_{n,m}=-\frac{\hbar e}{\mu \alpha^{2}}
  \left[
    \left(n+\frac{1}{2}+\frac{L}{2}\right)
    \frac{\omega_{c}}{\omega}-\frac{m-l}{2}
  \right].
\label{Mnm}
\end{equation}
In Fig. \ref{Magnetizacao}, we show the profile of the magnetization
as a function of the magnetic field for different values of
$\alpha$. We can see that profile of the oscillations changes with increasing magnetic field.
In the flat case, the magnetization presents both AB and dHvA-type
oscillations. The AB-type oscillations are due to the redistribution of the electron
states in the Fermi energy and are dominant in the weak magnetic fields
regime \ref{Magnetizacao-weak-B}.
\qquad
The dHvA-type oscillations become dominant over AB-type oscillations as
the magnetic field increases.
This is shown in Fig. \ref{MagnetizationDot3d} and they
are a result of the depopulation of a subband.
When only one subband is occupied in the Fermi energy, the AB-type
oscillations are absent (see Fig. \ref{Magnetization-1-subband}(d));
by comparison, we can note in Fig. \ref{Energy_all} that there is no
more state crossing, and therefore there are no more AB-type
oscillations.

For a quantum dot on a curved surface, we can also see a complex oscillation pattern.
As discussed above, the curvature effects tend to lift the degeneracy,
which directly result in the vanishing of the
magnetization at $B=0$. The AB-type oscillations are present, as we can see in
Fig. \ref{Magnetizacao-weak-B}(b). 
If only one subband is occupied in the Fermi energy, the AB-type
oscillations are absent (see Figs. \ref{Magnetization-1-subband}(a)-\ref{Magnetization-1-subband}(c)). Oscillations with larger amplitudes, which arise with the field raising, do not necessarily configure as dHvA-type oscillations. This can be seen in Figs. \ref{Magnetization-1-subband}(a)-\ref{Magnetization-1-subband}(c) where we see that the peak of the oscillations do not coincide with the depopulation of the subband with $n=1$.
Besides, in the interval where the lateral confinement is dominant, we
see that along with the AB-type oscillations, the magnetization presents
a peak.
This behavior is due to the fact that the $m=0$ state is not an
allowed state.
Figure \ref{MagnetizationDot3d} shows more clearly the region where the
AB-type oscillations are dominant.
Whenever the magnetic field is increased, we still have AB-type oscillations, but they have smaller
amplitudes.

\section{Persistent current}

The free energy of the isolated system allows us to extract another
important thermodymic quantity, namely, the persistent current.
By considering the variation of the magnetic flux confined to the hole
of the ring, the persistent current is calculated using the
Byers-Yang relation \cite{PRL.7.46.1961}
\begin{equation}
  I_{n,m}=-\frac{\partial E_{n,m}}{\partial \Phi}=
  - \frac{1}{\phi_{0}} \frac{\partial E_{n,m}}{\partial l},
\end{equation}
where $E_{n,m}$ is then given by Eq. (\ref{Enm_dot}) with $r_0=0$.
The total current is given by
\begin{equation}
I=\sum_{n,m}I_{n,m},  \label{Cor}
\end{equation}
where $I_{n,m}$ is explicitly given by
\begin{equation}
I_{n,m}=\frac{e\omega}{4\pi \alpha ^{2}}\left( \frac{m-l}{L}-\frac{\omega_{c}}{\omega}\right),  \label{Inm}
\end{equation}
which is the persistent current carried by a given state $\chi_{n,m}$ of
the ring.

\begin{figure}[t]
	\centering
	\includegraphics[width=\columnwidth]{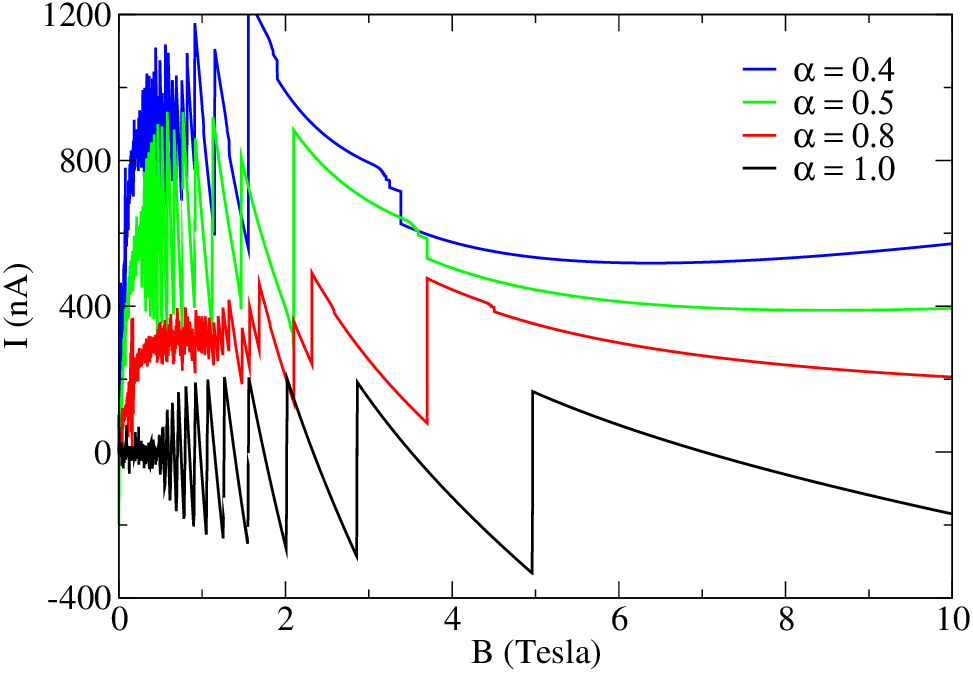}
	\caption{(Color online)
	The persistent currents of the quantum dot as a function of magnetic field
	strength.}
	\label{Corrente}
\end{figure}

\begin{figure}[t]
	\centering
	\includegraphics[width=\columnwidth]{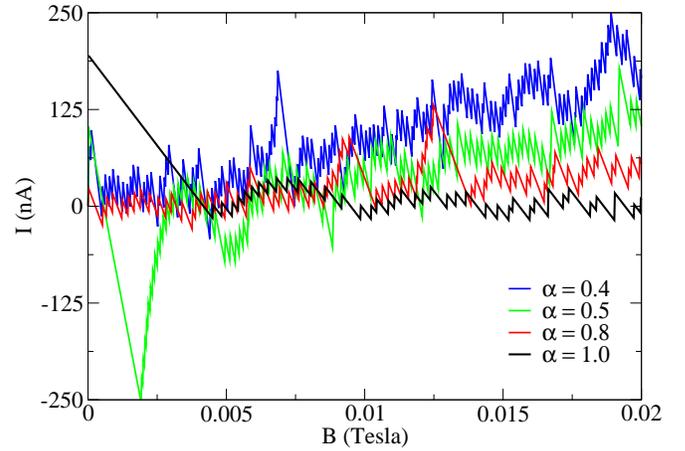}
	\caption{(Color online)
	The persistent currents in the weak magnetic fields interval.}
	\label{Corrente-weak-B}
\end{figure}

\begin{figure}[t!]
	\centering
	\includegraphics[width=\columnwidth]{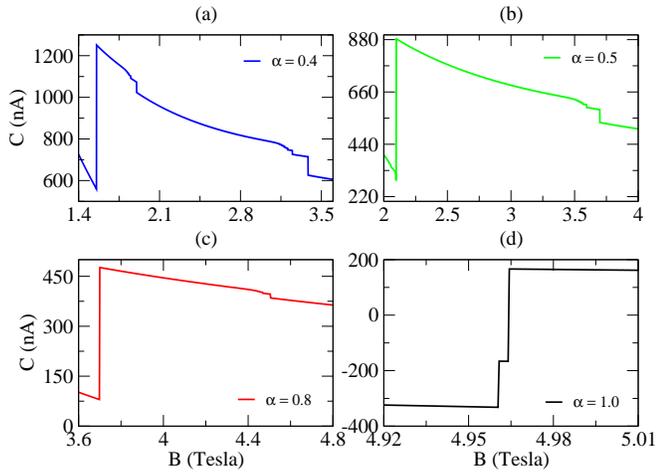}
	\caption{(Color online)The figure above shows the profile of the persistent currents when the subband with $n=1$ is depopulated.}
	\label{Corrente-1-subband}
\end{figure}

\begin{figure}[!h!]
  \centering
  \includegraphics[width=0.46\textwidth]{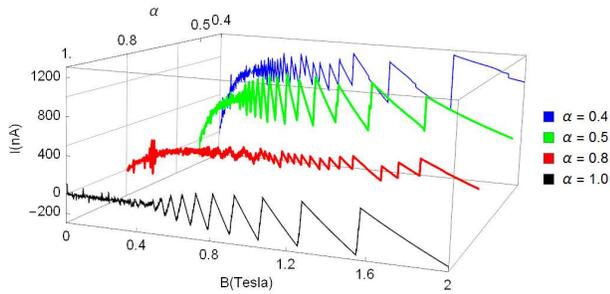}
  \caption{(Color online)
    The persistent currents of the quantum dot as a function of
    the magnetic field strength.}
  \label{PersistentCurrentDot3d}
\end{figure}

Remembering that in the case of a one-dimensional ring, the current is proportional to the magnetic moment. For the two-dimensional case, this relation is given by
\begin{equation}
M_{n,m}\left( B\right) =\pi r_{n,m}^{2}I_{n,m}-\frac{e\hbar}{\mu \alpha^{2}
}\left(n+\frac{1}{2}\right) \frac{\omega_{c}}{\omega}.
\label{Mnm_Inm}
\end{equation}
The above equation is a generalization of the classical result
between the current and the magnetic moment which is given by the first
term on the right side of the Eq. (\ref{Mnm_Inm}).
The second term results from the penetration of the magnetic field into
the $2$D structure, being it a diamagnetic term.
From this result it is possible to show that if
$\omega_{c}\ll \omega_{0}$ the persistent current and the magnetization
present a similar behavior.
Note that all the results above hold for the state of the quantum dot since
the wave function $\chi_{n,m}$ is zero in the $r=0$ region.
For $\alpha=1$, it was shown in Ref. \cite{PRB.1999.60.5626} that for
the $m=l$ state, the wave function is nonzero at $r=0$ and the Byers-Yang
relation no longer applies.
However, by using a appropriate limit, Tan and Inkson
\cite{PRB.1999.60.5626} calculated the persistent current carried by
this state.
In our case, we can write
\begin{equation}
I_{n,m=l}=\lim_{a_{1} \to 0}\lim_{m \to l}\lim_{\alpha \to 1} \frac{e\omega}{4\pi \alpha ^{2}}\left( \frac{m-l}{
        L}-\frac{\omega _{c}}{\omega}\right)=-\frac{e\omega_{c}}{4\pi},
\end{equation}
which is the current obtained in
Refs. \cite{PA.1993.200.504,PRB.1999.60.5626} for the $m-l=0$ state.
For $\alpha\neq1$, nevertheless, the state with $m=l$ does not
represent a problem due to the fact that it is not an allowed physical state.
The persistent current as a function of the magnetic field is plotted in
Fig. \ref{Corrente} for different values of $\alpha$.
On a flat surface, the energy levels of a quantum dot for the zero magnetic
field are highly degenerate, as it was already mentioned above.
This characteristic of the energy levels of a quantum dot influences
the behavior of the persistent current when the magnetic field is zero.
Note that the non-zero value of the persistent current at $B=0$ Tesla
(Fig. \ref{Corrente-weak-B}) makes it clear that the highest occupied
``shell" has not been fully filled.
A non-zero magnetic field redistributes the states so that more
states with $m>0$ are occupied for the same Fermi energy.
This gives rise to the AB-type oscillations
(Fig. \ref{Corrente-weak-B}).
However, by increasing the strength of the magnetic field, the dHvA-type oscillations
become more relevant (Figs. \ref{Corrente} and
\ref{PersistentCurrentDot3d}).
Notice that the current carried by a state with $m \leq 0$ is much
larger than that carried by a state with $m>0$.
In addition, the energies of states with $m \leq 0$ are always
near the bottom of a subband (See Fig. \ref{Corrente-1-subband}(d), for example) in the considered range of the magnetic field.
Therefore, when the bottom of the subband crosses the Fermi
energy, the depopulation of state with $m \leq 0$ occurs and
consequently it results in the abrupt change in the persistent
current \cite{PRB.1999.60.5626}.
Let us now analyze the quantum dot system when it is on a curved sample. We have seen above that the states of a quantum dot when $B= 0$ Tesla are two-fold degenerate due to the rotational symmetry. Because of this, the persistent current is zero. A small magnetic field breaks the rotational symmetry and the current shows the oscillations of AB-type as seen in Fig. \ref{Corrente-weak-B}. We can also observe that there is an increasing of the persistent current that occurs in the weak magnetic fields interval, which is due to the absence of the $m=0$ state. When the magnetic field increases, the oscillations that are observed do not necessarily configure dHvA-type oscillations. From the analysis of the persistent current in a flat sample, we know that the states with $m<0$ and $m=0$ have a great importance for the oscillations. In the case of the curved sample, the $m=0$ state is not allowed. Then, the oscillations that we observe result only from the depopulation of the states with $m<0$. As observed in the Fig. \ref{Energy_all}, these states are more affected by the curvature. Thus, we expect these oscillations to have a behavior different from that observed in a flat sample. In Figs. \ref{Corrente-1-subband}(a)-\ref{Corrente-1-subband}(c), we can observe more clearly that abrupt change occurs with the depopulation of a negative state, however these states are not necessarily near the bottom of a subband.

\section{Conclusions}
\label{Conc}

In this paper, we addressed the electronic properties, the
magnetization and the persistent current of a 2DEG in a quantum dot on a
conical surface and submitted to an external magnetic field. We have obtained analytically the wavefunctions and energy
eigenvalues of the model. It was shown that the energy is strongly influenced by the curvature of the surface which reflects the importance of the effect of the topology on such systems. It was verified that these changes are most significantly manifested
when the $\alpha$ parameter indicates a more pronounced curvature.
The oscillations of the AB that
appear in the profile of magnetization as a function of the magnetic
field for a quantum dot in the flat space remain when it is constrained
to a curved surface.
The AB-type oscillations are also observed in the analysis of the
persistent current. Nevertheless, the dHvA-type oscillations, which occur with a subband depopulation, are not well defined in both the persistent current and magnetization.
An anomalous behavior in the weak magnetic field
region in both the persistent current and the magnetization arose, and
we have found that this characteristic is a consequence of the absence
of the $m=0$ state. As a final word, we mention that  the magnetization of a 2DEG plays an important role in the context of a magnetically driven quantum heat engine \cite{PRE.2014.89.052107}.  So, the curvature effects addressed here could leave to the investigation concerned to the optimization  of such an engine in terms of work extraction and efficiency.  Moreover, the degeneracy has important consequences for this kind of quantum heat engine \cite{Entropy.2017.19.12}.  The considerations about this fact that we made here could impact it in a significant way.

\section*{Acknowledgments}
This work was partially supported by the Brazilian agencies CAPES, CNPq,
FAPEMA, FAPEMIG and Funda\c{c}\~{a}o Arauc\'{a}ria.

\bibliographystyle{apsrev4-2}

\end{document}